# Reliable Communications with Asymmetric Codebooks: An Information Theoretic Analysis of Robust Signal Hashing

Yücel Altuğ, M. Kıvanç Mıhçak *Member*, Onur Özyeşil, Vishal Monga

**Abstract**

In this paper, a generalization of the traditional point-to-point to communication setup, which is named as "reliable communications with asymmetric codebooks", is proposed. Under the assumption of independent identically distributed (i.i.d) encoder codewords, it is proven that the operational capacity of the system is equal to the information capacity of the system, which is given by $\max_{p(x)} I(U;Y)$, where $X, U$ and $Y$ denote the individual random elements of encoder codewords, decoder codewords and decoder inputs. The capacity result is derived in the "binary symmetric" case (which is an analogous formulation of the traditional "binary symmetric channel" for our case), as a function of the system parameters. A conceptually insightful inference is made by attributing the difference from the classical Shannon-type capacity of binary symmetric channel to the *gap* due to the codebook asymmetry.

## I. Introduction

We consider a point-to-point communication problem, where the codebooks of the encoder and the decoder are not the same, i.e., there is an asymmetry between them. In particular, we consider the scenario where the decoder's codebook is a perturbed version of the encoder's codebook; the statistical characterization of this perturbation is fixed and known by both the encoder and the decoder. Thus, the problem we consider can be viewed as a generalized version of the original reliable point-to-point communications problem stated by Shannon [1].

In our proposed setup, we aim to model a scenario, where the encoder (say, party A) and the decoder (say, party B) belong to collaborating, yet two different entities, which are communicating with each other: A would like to send a message to B. However, for the reasons of *privacy* (a clarifying practical example will be provided shortly), A does not want to share its codebook with B, while still maintaining a reliable communications link. Hence, due to the constraint of "reliable communications", B necessarily needs to possess a codebook which is "somehow related" to the codebook of A. In our case, we assume that, this "codebook relationship" is simply captured by a joint distribution of the codewords of both codebooks. In our formulation, we assume that the conditional distribution of the codewords of the codebook of B, conditioned on the codewords of the codebook of A, is fixed. Therefore, due to the Bayes' rule, the only design parameter is the marginal distribution of the codewords of the codebook of A. Our fundamental goal is to derive and characterize the maximum achievable rate of reliable information transmission between A and B in the aforementioned setup.

Y. Altuğ is with the School of Electrical and Computer Engineering, Cornell University, Ithaca, NY 14853, USA (e-mail: ya68@cornell.edu).

M. K. Mıhçak is with the Electrical and Electronic Engineering Department, Boğaziçi University, Istanbul, 34342, Turkey (e-mail: kivanc.mihcak@boun.edu.tr).

O. Özyeşil is with the Program in Applied and Computational Mathematics, Princeton University, Princeton, NJ, 08544, USA (e-mail: ozyesil@princeton.edu).

V. Monga is with Xerox Wilson Research Center, Webster, NY 14580, USA. (e-mail:vishal.monga@xeroxlabs.com)

Y. Altuğ was partially supported by TÜBİTAK Career Award no. 106E117; M. K. Mıhçak is partially supported by TÜBİTAK Career Award no. 106E117 and TÜBA-GEBIP Award.

A practical problem, which is closely related to the proposed setup, is *robust signal hashing* which is an area of research interest, particularly in signal processing and multimedia security. In the problem of robust signal hashing, the goal is to find a practical solution to "content tracking for anti-piracy search" with the aid of some side information at the receiver end. In that case, A is the owner of a valuable set of signals and would like to reliably find out whether any element of this set has been used without a proper consent; thus, A would like to keep track of such prohibited uses. Furthermore, usually the content owner A does not have the necessary resources to perform a desired anti-piracy search, and hence needs to utilize the resources of another entity (which accounts for B in the aforementioned setup). As a result, A would like to form a collaboration with B to carry out anti-piracy search, but at the same time does not want to reveal the "private" content itself due to its value (which accounts for the privacy issue in the aforementioned setup). Therefore, B should perform the anti-piracy search *only* with the aid of the side information provided by A about the original content. Here, the side information provided to the decoder (also termed as the receiver throughout the paper) side is termed as the "hash values" of the original content and the method via which they are constructed is termed as "the robust signal hashing algorithm" in robust signal hashing literature.

Adapting an information-theoretic approach to the robust signal hashing problem, we view the content owned by A as its codebook. The message transmission phase of the information-theoretic setup corresponds to publishing or broadcasting (without getting proper consent from A) the codewords of the encoder's codebook, possibly after introducing some disturbance[1]. The codebook made available to B (which is a "perturbed" version of the codebook of A) represents the " side information" (termed as the "hash values" of the valuable content in the multimedia security literature), using which the anti-piracy search is to be carried out. Note that, the statistical characterization of the perturbation between the codebooks of A and B represents the "robust signal hashing algorithm". As a result, the proposed problem of *reliable communications with asymmetric codebooks* constitutes the fundamental upper bound on the performance (i.e., the maximum rate of error-free information transmission) of any given robust signal hashing algorithm. We refer the interested reader to [7], [8], [9] for some practical robust signal hash algorithms proposed in the literature and [10] (resp. [11]) for a detection (resp. decision) theoretic treatment of the problem.

Next, we compare and contrast the reliable communications with asymmetric codebooks problem with the existing "related" formulations in the literature. First, observe that the problem at hand may be thought to belong to the class of traditional *side information* problems of Shannon theory (e.g. [3], [4], [5], [6]). However, this is indeed not the case due to the presence of asymmetric codebooks in our setup (which does not exist in the class of side information problems). For the case of "side information problems" dealing with channel coding, observe that the side information is about *the system parameters, and/or the noise corrupting the message*, but the codebooks employed in the system *always* shared between the parties, in other words, either the transmitter or the receiver is *favored* by the usage of the provided side information which is not available to the other party. However, for the proposed problem of reliable communications with asymmetric codebooks, there is *not* a shared codebook between the two communicating parties, (as hinted by the name of the proposed problem) and the system parameters (which amount to the statistical characterization of the system variables in our case, cf. Sec. II-B) are precisely known by both the encoder and the decoder. As a result, we believe that the proposed setup does not trivially reduce to the known classical problems of Shannon theory.

[1] The incorporation of a disturbance is modeled by the presence of a "noisy communication channel" in our problem, cf. Section II-B.

**Main Results:** For the proposed problem of reliable communications with asymmetric codebooks, we particularly focus on the scenario where:

(i) the alphabet, from which the encoder's codewords are drawn, is discrete and finite,

(ii) the communication channel between the encoder and the decoder is memoryless,

(iii) the statistical characterization of the perturbation between the codebooks (also termed as the "codebook perturbation" throughout the paper) of the encoder and the decoder is memoryless,

(iv) the codewords, which constitute the encoder codebook, are realizations of an i.i.d. (independent identically distributed) process[2].

Under these conditions, the main results of the paper are as follows:

(i) We derive the maximum rate of error-free information transmission (per communication channel use), termed as the "asymmetric codebook capacity" (cf. Theorem 3.1); it is shown to be the maximum of the mutual information between the decoder's codeword and the communication channel output, where the maximization is carried out over the probability distribution of the encoder's codeword.

(ii) We evaluate the asymmetric codebook capacity for a special case of interest (termed as "binary symmetric case"), where the encoder alphabet is binary, the codebook perturbation is a binary symmetric distribution, and the communication channel is a binary symmetric channel (BSC).

We begin our developments by stating the notation utilized in the paper and providing a rigorous statement of the problem formulation in Sec. II. In Section III, we state the main result of the paper: The forward and converse statements' proofs are given in Sections III-A and III-B, respectively; a closed form expression of the asymmetric codebook capacity for the binary symmetric case is presented in Sec. III-C. Paper ends with the concluding remarks in Sec. IV.

[2]The detailed justification of this assumption is given at the beginning of Sec. III.

## II. Notation and Problem Statement

### A. Notation

Boldface letters denote vectors; regular letters with subscripts denote individual elements of vectors. Furthermore, capital letters represent random variables and lowercase letters denote individual realizations of the corresponding random variable. The vector $[a_1, a_2, \ldots, a_N]^T$ is compactly represented by $\mathbf{a}^N$. The abbreviations "i.i.d.", "p.m.f.", and "w.l.o.g." are shorthands for the terms "independent identically distributed", "probability mass function", and "without loss of generality", respectively. For a discrete random variable $X$, with the corresponding p.m.f. denoted by $p\left(x\right)$ (where the subscript $X$ is omitted for simplicity, and should be evident from the context) defined on the alphabet $\mathcal{X}$, $H\left(X\right) = -\sum_{x\in\mathcal{X}} p\left(x\right)\log p\left(x\right)$ denotes its entropy[3]. Similarly, given discrete random variables $X$ and $Y$, the quantities $H\left(X,Y\right)$, $H\left(X|Y\right)$, $I\left(X;Y\right)$ denote the joint entropy of $X$ and $Y$, conditional entropy of $X$ given $Y$, and the mutual information between $X$ and $Y$, respectively. As a shorthand, *binary entropy* function is denoted by $H\left(p\right) \triangleq -p\log p - \left(1-p\right)\log\left(1-p\right)$ for $p \in [0,1]$.

### B. Problem Statement

In this section, we state the precise definition of the proposed problem of reliable communications with asymmetric codebooks. Such a communication system consists of two components: a *discrete-memoryless communication channel* denoted by $\left(\mathcal{X}, p\left(y|x\right), \mathcal{Y}\right)$ (with single letter input alphabet $\mathcal{X}$, single letter output alphabet $\mathcal{Y}$, single letter transition probability $p\left(y|x\right)$, cf. see [2], p. 193) and a $\left(2^{nR}, n\right)$ *asymmetric channel code* [4] (cf. Definition 2.1).

*Definition 2.1:* A $\left(2^{nR}, n\right)$ *asymmetric channel code* denoted by $\left(\mathcal{X}, \mathcal{C}_X, p\left(u|x\right), \mathcal{U}, \mathcal{C}_U\right)$ consists of the following components:

(i) a message set, $\mathcal{W} \triangleq \left\{1, \ldots, 2^{nR}\right\}$,

(ii) an encoder codebook, $\mathcal{C}_X \in \mathcal{X}^{2^{nR}\times n}$, consisting of length-$n$, $2^{nR}$ codewords, $\left\{\mathbf{x}^n\left(i\right)\right\}_{i=1}^{2^{nR}}$, each of which $j$-th element is denoted by $x_j\left(i\right)$, $1 \le i \le 2^{nR}$, $1 \le j \le n$.

(iii) an encoding function, $f : \mathcal{W} \to \mathcal{X}^n$, where $f\left(i\right) = \mathbf{x}^n\left(i\right)$,

(iv) a decoder codebook, $\mathcal{C}_U \in \mathcal{U}^{2^{nR}\times n}$, consisting of length-$n$, $2^{nR}$ codewords, $\left\{\mathbf{u}^n\left(i\right)\right\}_{i=1}^{2^{nR}}$, each of which $j$-th element is denoted by $u_j\left(i\right)$, $1 \le i \le 2^{nR}$, $1 \le j \le n$, such that $\mathcal{C}_U$ is formed from $\mathcal{C}_X$ via a probabilistic mapping (i.e., statistical perturbation) in a memoryless fashion, where

$$\Pr\left(\mathcal{C}_U|\mathcal{C}_X\right) = p\left(\mathbf{u}^n\left(1\right), \mathbf{u}^n\left(2\right), \ldots, \mathbf{u}^n\left(2^{nR}\right) | \mathbf{x}^n\left(1\right), \mathbf{x}^n\left(2\right), \ldots, \mathbf{x}^n\left(2^{nR}\right)\right) = \prod_{i=1}^{2^{nR}} p\left(\mathbf{u}^n\left(i\right)|\mathbf{x}^n\left(i\right)\right), \quad (1)$$

[3] Unless otherwise stated, all the logarithms are base-2.

[4] Throughout the paper, for the sake of convenience, we assume that $2^{nR} \in \mathbb{Z}^+$ for all $R \in \mathbb{R}^+ \cup \{0\}$ and for any $n \in \mathbb{Z}^+$.

and for all $1 \leq i \leq 2^{nR}$,

$$p\left(\mathbf{u}^n\left(i\right)|\mathbf{x}^n\left(i\right)\right) = \prod_{j=1}^{n} p\left(u|x\right)\Big|_{u=u_j(i),x=x_j(i)}, \tag{2}$$

(v) a decoding function, $g : \mathcal{Y}^n \rightarrow \mathcal{W} \cup \{0\}$, which is a deterministic mapping that assigns a decision (including a "null", denoted by $0$) to every received sequence $\mathbf{y}^n \in \mathcal{Y}^n$ via utilizing $\mathcal{C}_U$; the decoder output is denoted by $\hat{W}$,

(vi) a receiver side message-to-codebook mapping , $h : \mathcal{W} \rightarrow \mathcal{U}^n$, where $h\left(i\right) = \mathbf{u}^n\left(i\right)$ for all $1 \leq i \leq 2^{nR}$.

Next, we state several "error-event-related" definitions, which will be used throughout the paper. Note that, in all these events we condition on the particular realization of the decoder codebook $\mathcal{C}_U$ and explicitly state this in the notation.

- *Conditional probability of error, $\lambda_i$, conditioned on the transmitted message $i$ and the decoder codebook $\mathcal{C}_U$:*

$$\lambda_i\left(\mathcal{C}_U\right) \stackrel{\triangle}{=} \Pr\left(g\left(\mathbf{Y}^n\right) \neq i | h\left(i\right) = \mathbf{u}^n\left(i\right)\right) = \sum_{\mathbf{y}^n \in \mathcal{Y}^n} p\left(\mathbf{y}^n|\mathbf{u}^n\left(i\right)\right) 1_{(g(\mathbf{y}^n)\neq i)}, \tag{3}$$

where $1_{(\cdot)}$ is the standard indicator function.

- *Maximal probability of error, $\lambda^{(n)}$, conditioned on the decoder codebook $\mathcal{C}_U$:*

$$\lambda^{(n)}\left(\mathcal{C}_U\right) \stackrel{\triangle}{=} \max_{i \in \mathcal{W}} \lambda_i. \tag{4}$$

- *Average probability of error, $P_e^{(n)}$, conditioned on $\mathcal{C}_U$:*

$$P_e^{(n)} \stackrel{\triangle}{=} \Pr\left(W \neq g\left(\mathbf{Y}^n\right)\right). \tag{5}$$

The block diagram representation of the asymmetric communication system defined above is shown in Fig. 1 below.

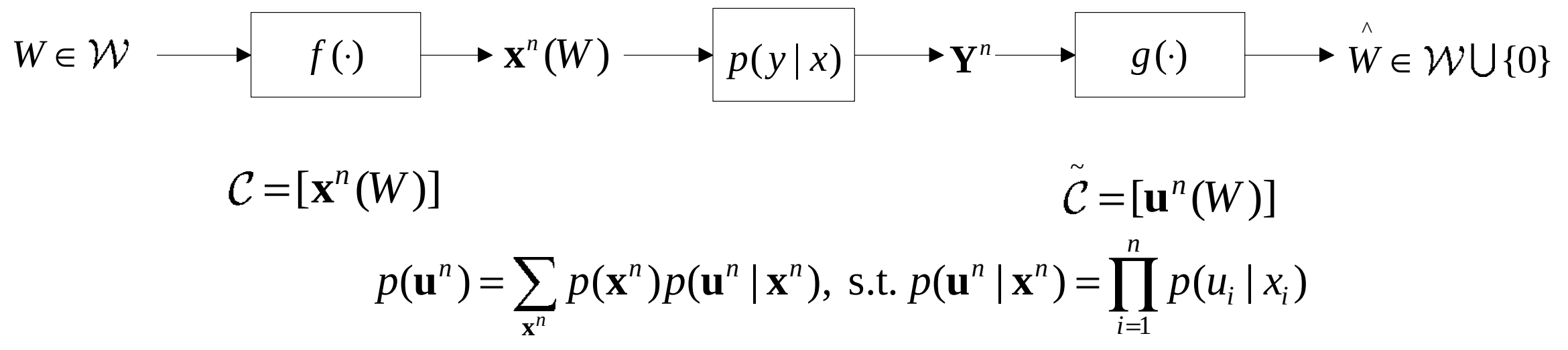

Fig. 1. The Block Diagram Representation of the Discrete Memoryless Channel with Asymmetric Codebooks.

*Remark 2.1:*

(i) The quantities $\mathcal{X}$, $\mathcal{U}$, $\mathcal{Y}$ represent the alphabets, from which the individual elements of the encoder codewords, the decoder codewords and the decoder inputs are drawn, respectively.

(ii) The p.m.f. $p\left(y|x\right)$ represents the (discrete memoryless) *communication channel* between the transmitter and the receiver, over which the transmission of information is carried out.

(iii) The p.m.f. $p\left(u|x\right)$ represents the *perturbation distribution* between the codewords of the encoder ($\left\{\mathbf{x}^n\left(i\right)\right\}_i$) and the decoder ($\left\{\mathbf{u}^n\left(i\right)\right\}_i$), which statistically characterizes the asymmetric nature of the codebooks of these components ($\mathcal{C}_X$ and $\mathcal{C}_U$).

(iv) Due to the physical nature of the problem, the generation of the receiver input and the generation of the receiver codebook, given the encoder codebook, are two separate independent events; in general, they do not necessarily need to happen at the same time, which highlights a fundamental difference between our setup and a "broadcast-channel-like" setup. Hence, given the random variables, $X \in \mathcal{X}$, $Y \in \mathcal{Y}$, $U \in \mathcal{U}$, obeying the conditional p.m.f.s $p\left(y|x\right)$ and $p\left(u|x\right)$, we have

$$p\left(y,u|x\right) = p\left(y|x\right)p\left(u|x\right), \tag{6}$$

which implies that $U \leftrightarrow X \leftrightarrow Y$ forms a Markov chain in the stated order.

(v) Both the encoder and the decoder possess the knowledge of $p\left(y|x\right)$, $p\left(u|x\right)$ and $p\left(x\right)$; on the other hand, the particular codebook realizations, $\mathcal{C}_X$ and $\mathcal{C}_U$ are known only by the encoder and the decoder, respectively.

(vi) The functions $f\left(\cdot\right)$ and $h\left(\cdot\right)$ effectively determine the "ordering" of the rows of the codebook matrices $\mathcal{C}_X$ and $\mathcal{C}_U$, respectively (w.l.o.g., the elements of the message set $\mathcal{W} = \left\{1, 2, \ldots, 2^{nR}\right\}$ are thought to be ordered in an increasing fashion at both sides). Since the encoder and the decoder *only* know the particular codebook realizations $\mathcal{C}_X$ and $\mathcal{C}_U$, respectively, equivalently they also *only* know the functions $f\left(\cdot\right)$ and $h\left(\cdot\right)$, respectively. Thus, the asymmetric nature of the problem arises from the mismatch between $f\left(\cdot\right)$ and $h\left(\cdot\right)$.

(vii) In this paper, the statistical perturbation that models the mapping from $\mathcal{C}_X$ to $\mathcal{C}_U$ is assumed to be *memoryless* for the sake of simplicity, which should be thought as a first step towards the direction of analyzing reliable communications with asymmetric codebooks. In the more general case, the mapping from $\mathcal{C}_X$ to $\mathcal{C}_U$ can be arbitrary, which constitutes part of our future research.

## III. Discrete-Memoryless Channel With Asymmetric Codebooks, I.I.D. Case

In this section, we proceed with analyzing the communication system presented in Section II-B under the following assumption: the codewords, $\left\{\mathbf{x}^n\left(i\right)\right\}_{i=1}^{2^{nR}}$, of the encoder's codebook, $\mathcal{C}_X$, are realizations of an i.i.d. random process with some marginal distribution $p(x)$. The main reason for this assumption is the fact that for the case of the dependent codewords of $\mathcal{C}_X$, we have dependency between the pairs $\left\{\left(u_i\left(W\right), Y_i\right)\right\}$ for all $W \in \mathcal{W}$ and $1 \leq i \leq n$ (cf. Lemma 3.1), which necessarily implies the existence of memory in the overall communication system[5]. Since we treat the current paper as a first step towards achieving the goal of investigating the problem of "reliable communications with asymmetric codebooks", we currently confine ourselves to the setup of i.i.d. encoder codewords for the sake of simplicity; analyzing more general cases, which include both asymmetric codebooks and memory, constitutes part of our future research.

Under the aforementioned assumptions of i.i.d. encoder codewords, memoryless perturbation distribution, and memoryless communications channel, we call the resulting system as *discrete memoryless communications channel with i.i.d. asymmetric*

[5]Note that as far as decoding is concerned, the overall setup may intuitively be thought to be analogous to a communication system, where both the encoder and the decoder share the same codebooks $\mathcal{C}_U$, and the communication channel via which the transmission of information is carried out is represented by the conditional p.m.f. $p\left(\mathbf{y}^n|\mathbf{u}^n\right)$. Note that the overall setup is *not* equivalent to the aforementioned communication system, since the design parameter in the setup of interest is $p(x)$. This issue will be further discussed in Remark 3.5.

*codebooks*[6]. Our fundamental result is Theorem 3.1 where we state the channel coding result for the resulting system. Section III-A contains achievability result, while Section III-B is devoted to the converse of Theorem 3.1. Section III concludes with the evaluation of the capacity for a special case, where the encoder codewords are binary, the communication channel is a binary symmetric channel and the perturbation distribution is a binary symmetric distribution, which is the topic of Section III-C.

*Definition 3.1:* An *i.i.d.* $\left(2^{nR}, n\right)$ *asymmetric channel code* for the discrete memoryless communications channel $\left(\mathcal{X}, p\left(y|x\right), \mathcal{Y}\right)$ consists of the six components mentioned in Definition 2.1 with the following additional property on the codewords of $\mathcal{C}_X$:

$$p\left(\mathbf{x}^n(w)\right) = \prod_{i=1}^{n} p\left(x_i(w)\right),$$

for all $w \in \mathcal{W}$, $1 \leq i \leq n$.

Now, we define the *achievable rate* and the operational capacity of the discrete memoryless channel with i.i.d. asymmetric codebooks.

*Definition 3.2:* A rate $R$ is said to be *achievable* provided that there exists a sequence of i.i.d. $\left(2^{nR}, n\right)$ asymmetric channel codes, such that the corresponding maximal probability of error (cf. (4)) $\lambda^{(n)}\left(\mathcal{C}_U\right) \to 0$ as $n \to \infty$.

*Definition 3.3:* For any given discrete memoryless channel $\left(\mathcal{X}, p\left(y|x\right), \mathcal{Y}\right)$, *the operational capacity* of the discrete memoryless channel with asymmetric codebooks, is defined as the supremum of the achievable rates.

Next, we define the information capacity of the discrete memoryless channel with i.i.d. asymmetric codebooks, which will be shown to be equal to the operational capacity of the system.

*Definition 3.4:* For any given discrete memoryless communications channel $\left(\mathcal{X}, p\left(y|x\right), \mathcal{Y}\right)$, *the information capacity* of the discrete memoryless channel with i.i.d. asymmetric codebooks, is defined as

$$C \triangleq \max_{p(x)} I(U;Y), \tag{7}$$

where $p(u,y) \triangleq \sum_{x \in \mathcal{X}} p(x)p(y|x)p(u|x)$ (cf. (6)).

*Remark 3.1:* We note two observations: First, $I\left(U;Y\right) \geq 0$ is upper-bounded by some finite value since both $\mathcal{U}$ and $\mathcal{Y}$ are discrete finite sets. Second, the set of all probability vectors $p\left(x\right)$ belong to a closed and bounded subset of $[0,1]^{|\mathcal{X}|}$. Combining these observations, we deduce that the maximum in (7) exists.

*Remark 3.2:* Note that, the quantity $I\left(U;Y\right)$, the argument of (7), is also the argument of the maximization problem corresponding to the classical channel capacity expression

$$\max_{p(u)} I\left(U;Y\right), \tag{8}$$

[6]We show in Lemma 3.2 that, under the specified assumptions, the elements of the decoder codebook are also i.i.d.

where $U$ and $Y$ are the communication channel input and the corresponding channel output, respectively. While the arguments of (7) and (8) are the same, the optimization parameter is different ( $p\left(x\right)$ and $p\left(u\right)$ in the former and latter, respectively). Thus, relating our problem to the classical channel capacity problem, the result (7) is quite intuitive: The argument $I\left(U;Y\right)$ represents the amount of information that can be transmitted through the channel since $U$ and $Y$ denote the decoder's codebook random variable and the communication channel output random variable, respectively; on the other hand, the maximization is carried out over $p\left(x\right)$, the distribution of the encoder codebook random variable, which is the only design parameter.

*Theorem 3.1:* (Channel Coding Theorem With Asymmetric Codebooks) For a discrete memoryless channel with i.i.d. asymmetric codebooks, all rates below capacity $C$ are achievable. Specifically, for every rate $R < C$, there exists a sequence of $\left(2^{nR}, n\right)$ i.i.d. asymmetric channel codes with maximum probability of error can be made arbitrarily small for sufficiently large $n$.

Conversely, any sequence of $\left(2^{nR}, n\right)$ i.i.d. asymmetric codes with asymptotically vanishing maximum error probability should necessarily satisfy $R \leq C$.

*Remark 3.3:* Using the chain rule for mutual information, we have

$$I\left(U,X;Y\right) = I\left(U;Y\right) + I\left(X;Y|U\right), \tag{9}$$

$$= I\left(X;Y\right) + I\left(U;Y|X\right), \tag{10}$$

Combining (9), (10) and noting that $I\left(U;Y|X\right) = 0$ (cf. (6)), we get

$$I\left(U;Y\right) = I\left(X;Y\right) - I\left(X;Y|U\right). \tag{11}$$

Recall that, if there is no perturbation between the encoder and decoder codebooks (i.e., $\mathcal{C}_U = \mathcal{C}_X$), the proposed system reduces to the conventional channel coding setup, in which case (given $p\left(x\right)$) the achievable rate is $I\left(X;Y\right)$. Hence, inspecting (11), we observe that the term $I\left(X;Y|U\right) \geq 0$ can be viewed as the *achievable rate loss due to the asymmetric codebooks*.

Furthermore, $I\left(X;Y|U\right) = 0$ if and only if $p\left(x,y|u\right) = p\left(x|u\right)p\left(y|u\right)$, i.e. $X \leftrightarrow U \leftrightarrow Y$ forms a Markov chain. Since we also have $U \leftrightarrow X \leftrightarrow Y$ (cf. (6)), this is possible if and only if there exists a one-to-one mapping between $U$ and $X$.

Next, we state following lemma, which states that in case of i.i.d. encoder codewords, the pairs, consisting of the individual elements of the decoder codewords and the corresponding communication channel outputs, are i.i.d. This result shall be used in proving the achievability and converse theorems.

*Lemma 3.1:* Given $p\left(\mathbf{x}^n\right) = \prod_{i=1}^{n} p\left(x_i\right)$, a discrete memoryless communications channel $\left(\mathcal{X}, p\left(y|x\right), \mathcal{Y}\right)$, and a $\left(2^{nR}, n\right)$ asymmetric channel code $\left(\mathcal{X}, \mathcal{C}_X, p\left(u|x\right), \mathcal{U}, \mathcal{C}_U\right)$, we have

$$p\left(\mathbf{y}^n, \mathbf{u}^n\right) = \prod_{i=1}^{n} p\left(y_i, u_i\right), \tag{12}$$

where $p\left(y,u\right)=\sum_{i=1}^{n}p(x)p(y|x)p(u|x)$.

*Proof:* See Appendix I. ■

### A. Achievability

*Theorem 3.2:* (Achievability) For every rate $R<C$, there exists a sequence of $(2^{nR},n)$ i.i.d. asymmetric codes with arbitrarily small maximum probability of error for sufficiently large $n$.

*Proof:* The proof relies on the random coding arguments. First, we state the achievable rate for any given $p\left(x\right)$.

**Encoding:**

(i) Generation of Codebooks: Fix $p(x)$ and reveal to both sides. Generate the encoder codebook $\mathcal{C}_X$ as stated in Definition 2.1, part (ii), such that $x_i(w)$ are i.i.d. realizations of $X$ of which distribution is $p(x)$ for all $i\in\{1,\ldots,n\}$, $w\in\mathcal{W}$. Construct the decoder's codebook $\mathcal{C}_U$, as stated in Definition 2.1, part (iv), using the conditional p.m.f. $p\left(u|x\right)$.

(ii) Choose a message $w$ uniformly from $\mathcal{W}$ (i.e., $\Pr\left(W=w\right)=2^{-nR}$ for all $w\in\mathcal{W}$). Then $f\left(w\right)=\mathbf{x}^n(w)$ is transmitted over the communication channel $p(y|x)$, resulting in $\mathbf{Y}^n$, such that $\Pr\left(\mathbf{Y}^n=\mathbf{y}^n|\mathbf{x}^n\left(w\right)\right)=\prod_{i=1}^{n}p\left(y_i|x_i\left(w\right)\right)$ (recall the memoryless property of the communication channel).

**Decoding:**

(i) Note that, $\left\{u_i\left(W\right),Y_i\right\}_{i=1}^{n}$ pairs are independent of each other (cf. Lemma 3.1), where $\mathbf{Y}^n$ is the communication channel output corresponding to the message $W\in\mathcal{W}$. Next, we use jointly typical decoding: Decide the unique $\hat{W}\in\mathcal{W}$ (if exists), such that $\left(\mathbf{u}^n\left(\hat{W}\right),\mathbf{Y}^n\right)\in A_\epsilon^{(n)}\left(U,Y\right)$, where $A_\epsilon^{(n)}\left(U,Y\right)$ (from now on denoted by $A_\epsilon^{(n)}$ for the sake of simplicity) is the $\epsilon$-jointly-typical set [2], defined on $p\left(u,y\right)=\sum_{x\in\mathcal{X}}p\left(x\right)p\left(y|x\right)p\left(u|x\right)$:

$$A_\epsilon^{(n)}\triangleq\left\{\left(\mathbf{u}^n,\mathbf{y}^n\right)\ :\ |-\frac{1}{n}\log p\left(\mathbf{u}^n\right)-H(U)|<\epsilon,\ |-\frac{1}{n}\log p\left(\mathbf{y}^n\right)-H(Y)|<\epsilon,\ |-\frac{1}{n}\log p\left(\mathbf{u}^n,\mathbf{y}^n\right)-H(U,Y)|<\epsilon\right\}. \tag{13}$$

If such a $\hat{W}\in\mathcal{W}$ is not unique or does not exist, then declare $g\left(\mathbf{Y}^n\right)=0$. The error event is defined as

$$\mathcal{E}\triangleq\left\{\hat{W}\neq W\right\}. \tag{14}$$

**Analysis of the Probability of Error:**

Observe that, using the uniform distribution of $W$ over $\mathcal{W}$, we have

$$P_e^{(n)}=\sum_{i=1}^{2^{nR}}\Pr\left(g\left(\mathbf{Y}^n\right)\neq i|h\left(i\right)=\mathbf{u}^n\left(i\right)\right)\Pr\left(W=i\right)=\frac{1}{2^{nR}}\sum_{i=1}^{2^{nR}}\lambda_i. \tag{15}$$

Next, we show that the elements of the codewords of the decoder codebook are i.i.d.

*Lemma 3.2:*

$$\Pr\left(\mathcal{C}_U\right) = \prod_{i=1}^{n} \prod_{w=1}^{2^{nR}} p(u_i(w)), \tag{16}$$

where $p(u) \triangleq \sum_x p(x)p(u|x)$.

*Proof:* See Appendix II. ■

Recalling the definitions (14) and (5), and using (15), we have the following average probability of error, averaged over all possible decoder codebooks[7]:

$$\begin{aligned} P_e^{(n)} &= \Pr\left(\mathcal{E}\right), \\ &= \sum_{\mathcal{C}_U} \Pr\left(\mathcal{C}_U\right) P_e^{(n)}\left(\mathcal{C}_U\right), \\ &= \frac{1}{2^{nR}} \sum_{w=1}^{2^{nR}} \sum_{\mathcal{C}_U} \Pr\left(\mathcal{C}_U\right) \lambda_w\left(\mathcal{C}_U\right), \\ &= \sum_{\mathcal{C}_U} \Pr\left(\mathcal{C}_U\right) \lambda_1\left(\mathcal{C}_U\right), \qquad (17) \\ &= \Pr\left(\mathcal{E}|W=1\right), \qquad (18) \end{aligned}$$

where (17) follows since the decoder's codebook generation is symmetric per Lemma 3.2. Thus, w.l.o.g., from now on we confine ourselves to the case of $W = 1$.

Next, we define the following events of joint typicality of $\left(\mathbf{u}^n(i), \mathbf{Y}^n\right)$ in case of $W = 1$:

$$\mathcal{E}_i \triangleq \left\{ \left(\mathbf{u}^n(i), \mathbf{Y}^n\right) \in A_\epsilon^{(n)} \,\middle|\, W = 1 \right\}, \tag{19}$$

for $i \in \left\{1, \ldots, 2^{nR}\right\}$.

As a result, we have

$$\begin{aligned} P_e^{(n)} &= \Pr\left(\mathcal{E}|W=1\right), \qquad (20) \\ &= \Pr\left(\mathcal{E}_1^c \cup \bigcup_{j=2}^{2^{nR}} \mathcal{E}_j | W=1\right), \qquad (21) \\ &\leq \Pr\left(\mathcal{E}_1^c|W=1\right) + \sum_{j=2}^{2^{nR}} \Pr\left(\mathcal{E}_j|W=1\right). \qquad (22) \end{aligned}$$

where (20) follows from (18), (21) follows from the definition (19), (22) follows from the standard union bound. Next, we provide upper bounds for the terms constituting the right hand side of (22):

$$\Pr\left(\mathcal{E}_1^c|W=1\right) \leq \epsilon, \tag{23}$$

$$\Pr\left(\mathcal{E}_j|W=1\right) \leq 2^{-n(I(U;Y)-3\epsilon)}, \quad \text{for any } j \in \left\{2, 3, \ldots, 2^{nR}\right\} \tag{24}$$

[7]Recall that since the marginal probability of $\mathcal{C}_U$ is the result of averaging the joint distribution of $\mathcal{C}_X$ and $\mathcal{C}_U$ over $\mathcal{C}_X$, (cf. Lemma 3.2) we equivalently average out the conditional probability of error expression over the two codebooks of the system, since the probability space of $\mathcal{C}_U$ is jointly induced by the perturbation distribution and the probability space of $X$. This point constitutes the fundamental difference between the achievability proof for our system and the classical achievability proof of channel coding.

for any $\epsilon > 0$ and sufficiently large $n$; here (23) and (24) follow from the joint AEP theorem (cf. Theorem 7.6.1. of [2]) since $\mathbf{u}^n(i)$ and $\mathbf{u}^n(1)$ are independent for $i \neq 1$ (cf. Lemma 3.2). Using (23) and (24) in (22) yields,

$$\begin{aligned} P_e^{(n)} &\leq \epsilon + \sum_{j=2}^{2^{nR}} 2^{-n(I(U;Y)-3\epsilon)}, \\ &\leq \epsilon + 2^{nR} 2^{-n(I(U;Y)-3\epsilon)}, \\ &= \epsilon + 2^{-n(I(U;Y)-R-3\epsilon)}, \end{aligned} \tag{25}$$

for any $\epsilon > 0$ and sufficiently large $n$. Note that, (25) implies that if $I(U;Y) - R > 3\epsilon$, we have

$$P_e^{(n)} < 2\epsilon, \tag{26}$$

for sufficiently large $n$. Thus, for any rate $R < I(U;Y)$, there exists a sufficiently small $\epsilon$ and a sufficiently large $n$, such that (26) holds. Next, choose $p(x)$ in the encoding step so as to maximize $I(U;Y)$; let $p^*(x)$ be a maximizer. Then the condition $R < I(U;Y)$ can be replaced by the achievability condition $R < C$. Next, following similar steps to those used in the achievability proof of the classical channel coding theorem (cf. [2], p. 204), we conclude that we can construct a code of rate $R - 1/n$ with maximal probability of error $\lambda^{(n)} < 4\epsilon$ for any $\epsilon > 0$ for sufficiently large $n$. ■

*Remark 3.4:* Note that, since the decoder knows the marginal distribution of the encoder's codewords, it "typically" knows (due to the AEP) the codewords of the encoder as a "cluster". Furthermore, due to the availability of the statistical characterization of the communication channel to both sides, the decoder also knows the jointly typical $(\mathbf{x}^n, \mathbf{y}^n)$ pairs, again as a "cluster". However, since the decoder does not know the precise "ordering"[8] of the encoder's codebook (which uniquely determines $f(\cdot)$), the clustering information by itself does not yield anything useful as far as decoding is concerned: The decoder may very well find out the particular codeword $\mathbf{x}^n$ which is jointly typical with the received $\mathbf{y}^n$; however in the absence of the encoding function $f(\cdot)$, it can not calculate $f^{-1}(\mathbf{x}^n)$ in order to give an estimate of the transmitted message. Hence, the only tool decoder may use in order to perform detection is the codebook, $\mathcal{C}_U$, made available via the usage of the perturbation distribution, and the resulting function $h(\cdot)$.

### *B. Converse*

In this section, we provide the converse of Theorem 3.1, which is stated below:

*Theorem 3.3:* (Converse) For any $\left(2^{nR}, n\right)$ i.i.d. asymmetric codes with $\lambda^{(n)} \to 0$, we have $R < C$.

*Proof:* The proof consists of three steps:

Step 1: We first show that, for $\left(2^{nR}, n\right)$ asymmetric codes with $\lambda^{(n)} \to 0$, we necessarily need to have that $h(\cdot)$ is one-to-one (cf. Lemma 3.3).

[8]Recall that the encoder codebook, $\mathcal{C}_X$ is not available at the decoder.

Step 2: Then, we show that, for $\left(2^{nR}, n\right)$ i.i.d. asymmetric codes with $P_e^{(n)} \to 0$ and $h(\cdot)$ one-to-one, we necessarily have $R < C$.

Step 3: Next, we note that every $\left(2^{nR}, n\right)$ i.i.d. asymmetric code with $\lambda^{(n)} \to 0$ should also necessarily satisfy $P_e^{(n)} \to 0$. Since such codes also must satisfy the property of $h(\cdot)$ being one-to-one (per Step 1 above, by recalling that the set of i.i.d. asymmetric codes is a subset of asymmetric codes), these codes satisfy the conditions specified in Step 2. Thus, proving the statement of Step 2 above constitutes a sufficient condition for the converse theorem.

We proceed with proving the following lemma which establishes Step 1 above.

*Lemma 3.3:* For any $\left(2^{nR}, n\right)$ asymmetric code with $\lambda^{(n)} \to 0$, $h(\cdot)$ is necessarily a one-to-one mapping.

*Proof:* See Appendix III. ■

Next, we continue with completing the second step of the proof: We show that, for $\left(2^{nR}, n\right)$ i.i.d. asymmetric codes with $P_e^{(n)} \to 0$ and $h(\cdot)$ one-to-one, we necessarily have $R < C$.

First note that, the transmitted message, $W$, and the communication channel output, $\mathbf{Y}^n$, have a joint distribution; and that the decoder output, $\hat{W}$ is a function of the communication channel output, $\mathbf{Y}^n$. Hence, we conclude that $W \leftrightarrow \mathbf{Y}^n \leftrightarrow \hat{W}$ forms a Markov chain in the specified order. Furthermore, since $h(W) = \mathbf{u}^n(W)$ is a function of $W$, we also see that $\mathbf{u}^n(W) \leftrightarrow W \leftrightarrow \mathbf{Y}^n$ forms a Markov chain in the specified order. Combining these two observations, we conclude that $\mathbf{u}^n(W) \leftrightarrow W \leftrightarrow \mathbf{Y}^n \leftrightarrow \hat{W}$ forms a Markov chain in the specified order. Next, since $h(\cdot)$ is one-to-one per assumption, the previous Markov chain further implies that $W \leftrightarrow \mathbf{U}^n \leftrightarrow \mathbf{Y}^n \leftrightarrow \hat{W}$ forms a Markov chain in the specified order, where $\mathbf{U}^n$ denotes $\mathbf{u}^n(W)$ for the sake of simplicity.

Now, we continue with investigating the pair $\mathbf{U}^n, \mathbf{Y}^n$ in the following chain of equalities:

$$\begin{aligned} p\left(\mathbf{y}^n|\mathbf{u}^n\right) &= \frac{p\left(\mathbf{y}^n, \mathbf{u}^n\right)}{p\left(\mathbf{u}^n\right)}, \\ &= \frac{\prod_{i=1}^n p\left(y_i, u_i\right)}{p\left(\mathbf{u}^n\right)}, \qquad (27) \\ &= \frac{\prod_{i=1}^n p\left(y_i, u_i\right)}{\prod_{i=1}^n \left[\sum_{x_i} p(u_i|x_i)p(x_i)\right]}, \qquad (28) \\ &= \prod_{i=1}^n p(y_i|u_i), \qquad (29) \end{aligned}$$

where (27) follows from (12), (28) follows from (2) and the fact that $p\left(\mathbf{x}^n\right) = \prod_{i=1}^n p(x_i)$ and (29) follows from recalling that $p(y,u) = \sum_{x_i} p(y|x)p(u|x)p(x)$ and noting that $p(y_i|u_i) = \frac{p(y_i,u_i)}{\sum_{x_i} p(u_i|x_i)p(x_i)}$ due to Bayes rule. Equipped with the memoryless property of $p\left(\mathbf{y}^n|\mathbf{u}^n\right)$ (cf. (29)), we continue with the following chain of inequalities:

$$\begin{aligned} nR &= H(W), \qquad (30) \\ &= I(\hat{W}; W) + H(W|\hat{W}), \end{aligned}$$

$$\leq \quad I(\mathbf{U}^n;\mathbf{Y}^n) + \left(1 + nRP_e^{(n)}\right), \tag{31}$$

$$= \quad H\left(\mathbf{Y}^n\right) - \sum_{i=1}^{n} H\left(Y_i|U_i\right) + \left(1 + nRP_e^{(n)}\right), \tag{32}$$

$$= \quad \left(1 + nRP_e^{(n)}\right) + \sum_{i=1}^{n} \left(H(Y_i) - H(Y_i|U_i)\right), \tag{33}$$

$$= \quad \left(1 + nRP_e^{(n)}\right) + \sum_{i=1}^{n} I(U_i;Y_i), \tag{34}$$

where (30) follows since $W$ is uniformly distributed over $\mathcal{W}$, (31) follows using Fano's inequality (the second term) and the data processing inequality [2] (the first term) by recalling that $W \leftrightarrow \mathbf{U}^n \leftrightarrow \mathbf{Y}^n \leftrightarrow \hat{W}$ forms a Markov chain in the specified order, (32) follows using (29), (33) follows since $p\left(\mathbf{y}^n\right) = \prod_{i=1}^{n}\sum_{x_i} p(y_i|x_i)p(x_i) = \prod_{i=1}^{n} p\left(y_i\right)$ (because of the fact that communication channel is memoryless and $p\left(\mathbf{x}^n\right) = \prod_{i=1}^{n} p(x_i)$), which implies that $H\left(\mathbf{Y}^n\right) = \sum_{i=1}^{n} H(Y_i)$, and (34) follows using the definition of mutual information.

Next, we aim to upper bound the second term of (34). We proceed with noting that

$$I(U;Y) \quad = \quad \sum_{u,y} p(u,y) \log \frac{p(y|u)}{p(y)},$$

$$= \quad \sum_{u,y}\sum_{x} p(x)p(u|x)p(y|x) \log \frac{\frac{\sum_x p(x)p(y|x)p(u|x)}{\sum_x p(x)p(u|x)}}{\sum_x p(x)p(y|x)}, \tag{35}$$

where (35) follows using (6). Note that, the only variable that can be "adjusted" in (35) is $p\left(x\right)$ (the variables $p\left(y|x\right)$ and $p\left(u|x\right)$ are given and fixed); hence, we conclude that

$$I(U_i;Y_i) \leq \max_{p(x)} I(U;Y) = C, \text{ for all } (U_i,Y_i) \text{ pairs.} \tag{36}$$

Using (36) in (34), we have

$$R \quad \leq \quad \frac{1}{n} + RP_e^{(n)} + C,$$

$$\leq \quad \epsilon + C, \tag{37}$$

for any $\epsilon > 0$ and sufficiently large $n$, where (37) follows since $P_e^{(n)} \to 0$ and $1/n \leq \epsilon$ for sufficiently large $n$. Therefore, (37) implies $R < C$, which concludes the proof, since it is a sufficient condition for the validity of the statement: 'for any i.i.d. asymmetric code with $\lambda^{(n)} \to 0$, we necessarily have $R < C$', per Step 3. ■

*Remark 3.5:*

(i) Note that the fact "$W \leftrightarrow \mathbf{U}^n \leftrightarrow \mathbf{Y}^n \leftrightarrow \hat{W}$ forms a Markov chain", also has an intuitive explanation: From the receiver side, the *only* channel (a hypothetical channel, depends on the choice of $p(x)$) between the encoder and the decoder is $p(\mathbf{y}^n|\mathbf{u}^n)$ since there's a deterministic relation between $W, \mathbf{x}^n(W), \mathbf{u}^n(W)$, recalling the formation of the decoder's codebook[9]. Therefore, although the transmitter sends $\mathbf{x}^n(w)$ using the original communication channel $p(y|x)$, the effective situation from the receiver side is as follows: The transmitter sends $\mathbf{u}^n(W)$ using the hypothetical channel $p(\mathbf{y}^n|\mathbf{u}^n)$ and the channel $p(\mathbf{y}^n|\mathbf{u}^n)$ is *known* at the decoder's side (since $p(x)$, $p(y|x)$ and $p(u|x)$ are all known by the decoder); therefore, the resulting problem at hand is analogous (although not equivalent) to the classical point-to-point communication variant mentioned in Remark 3.2 (the difference being the argument over which the maximization is carried out).

(ii) The meaning of the hypothetical channel $p(\mathbf{y}^n|u^n)$ mentioned in item (i) of this remark can be explained as follows: Since the received $\mathbf{Y}^n$ is due to sending $\mathbf{x}^n(W)$ through the channel $p(y|x)$, and the corresponding $\mathbf{u}^n(W)$ is produced via perturbing $\mathbf{x}^n(W)$, we observe that the dependence of $\mathbf{Y}^n$ on $\mathbf{u}^n(W)$ is over $\mathbf{x}^n(W)$. Therefore, the decoder (in some sense) "derives an estimate of" $\mathbf{x}^n(W)$ first (through the usage of $p(\mathbf{x}^n|\mathbf{u}^n)$, which can be evaluated at the receiver side, since $p(x)$ and $p(u|x)$ are available at the decoder), and then decides on $\hat{W}$ using $p(y|x)$ and the aforementioned "derived estimate of" $\mathbf{x}^n(W)$.

### *C. Binary Symmetric Case*

In this section, we consider a specific example of the discrete memoryless channel with i.i.d. asymmetric codebooks, which is shown in Figure 1. In particular, we consider the case for which $\mathcal{X} = \mathcal{Y} = \mathcal{U} = \{0, 1\}$, the communication channel (resp. the perturbation distribution) is the binary symmetric channel (resp. the binary symmetric distribution) with crossover probability $p_1$ (resp. $p_2$). Thus, we have

$$Y = X \oplus Z_1, \tag{38}$$

where $\Pr(Z_1 = 1) = p_1$ and $\Pr(Z_1 = 0) = 1 - p_1$;

$$U = X \oplus Z_2, \tag{39}$$

where $\Pr(Z_2 = 1) = p_2$ and $\Pr(Z_2 = 0) = 1 - p_2$, where $\oplus$ denotes addition modulo 2.

*Theorem 3.4:* For the binary symmetric case of the discrete codebook channel with i.i.d. asymmetric codebooks, the capacity is given by

$$C = 1 - H(p_1 + p_2(1 - 2p_1)), \tag{40}$$

where capacity is achieved if and only if $X$ is a Bernoulli $1/2$ random variable, i.e., $Pr(X = 1) = Pr(X = 0) = 1/2$.

*Proof:* First of all, we define the auxiliary random variable $V$ as follows:

$$V \triangleq X \oplus Z_1 \oplus Z_2. \tag{41}$$

[9]Note that this deterministic relation *does not* contradict with the existence of the probabilistic mapping, i.e. perturbation distribution, in the formation of the decoder's codebook. This deterministic mapping points out the relation between $W$, $\mathbf{x}^n(W)$ and $\mathbf{u}^n(W)$ *after* the formation of both $\mathcal{C}_X$ and $\mathcal{C}_U$.

Next, we provide the following lemma:

*Lemma 3.4:* $U \leftrightarrow V \leftrightarrow Y$ forms a Markov chain, i.e.

$$p(u, y|v) = p(u|v)p(y|v). \tag{42}$$

*Proof:* See Appendix IV. ∎

Next, observe that from the definition of auxiliary random variable, $V$ (cf. (41)), we see that $X \leftrightarrow U \leftrightarrow V$ forms a Markov chain in the specified order, i.e.,

$$p(x, v|u) = p(x|u)p(v|u), \tag{43}$$

and $X \leftrightarrow Y \leftrightarrow V$ forms a Markov chain in the specified order, i.e.,

$$p(x, v|y) = p(x|y)p(v|y). \tag{44}$$

Furthermore, combining (6) and (42) yields the following "circular Markov chain" structure between $X$, $Y$, $V$, $U$, which helps to visualize the situation at hand better (such a construction seems to be new to the best of our knowledge):

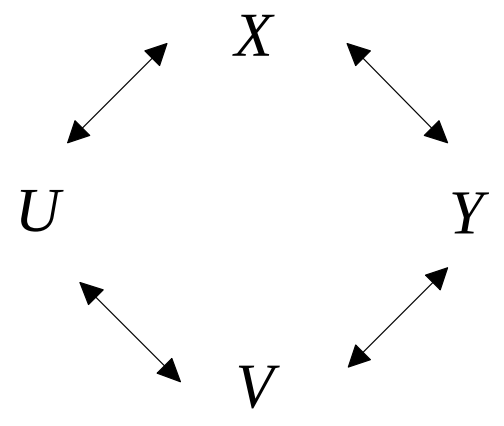

Fig. 2. The *circular Markov chain structure* of the random variables $X$, $Y$, $V$, $U$, defined for the binary symmetric communication channel and the perturbation distribution of Section III-C.

Using the definition of mutual information, we have

$$\begin{aligned} I(V;U) - I(V;U\,|\,Y) &= H(U) - H(U\,|\,V) - H(U\,|\,Y) + H(U\,|\,V,Y), \\ &= I(U;Y) - I(U;Y\,|V), \\ &= I(U;Y), \end{aligned} \tag{45}$$

where (45) follows since $U \leftrightarrow V \leftrightarrow Y$ forms a Markov chain in the specified order (cf. (42)). Similarly, using the definition of mutual information, we have

$$\begin{aligned} I(V;Y) - I(V;Y\,|\,U) &= H(V) - H(V\,|\,Y) - H(V\,|\,U) + H(V\,|\,U,Y), \\ &= H(V) + H(V\,|\,U,Y) - H(p_1) - H(p_2), \end{aligned} \tag{46}$$

where (46) follows using (38), (39) and (41). Noting that,

$$I\left(U,Y;V\right) = I\left(V;U\right) + I\left(V;Y\,|\,U\right), \tag{47}$$

$$= I\left(V;Y\right) + I\left(V;U\,|\,Y\right), \tag{48}$$

due to the chain rule for mutual information, we get

$$I\left(V;U\right) - I\left(V;U\,|\,Y\right) = I\left(V;Y\right) - I\left(V;Y\,|\,U\right). \tag{49}$$

Using (45) and (46) in the left hand side and the right hand side of (49), respectively, we have

$$I\left(U;Y\right) = H\left(V\right) + H\left(V\,|\,U,Y\right) - H\left(p_1\right) - H\left(p_2\right). \tag{50}$$

Next, we proceed with evaluating $H\left(V\,|\,U,Y\right)$. Using the chain rule for entropy, we have

$$H\left(X,U,Y,V\right) = H\left(U,Y\,|\,X,V\right) + H\left(X,V\right), \tag{51}$$

$$= H\left(X,V\,|\,U,Y\right) + H\left(U,Y\right), \tag{52}$$

Evaluating the individual terms in (51) and (52), we get

$$H\left(U,Y\,|\,X,V\right) = H\left(U\,|\,X,V\right) + H\left(Y\,|\,X,V\right), \tag{53}$$

$$H\left(X,V\right) = H\left(V\,|\,X\right) + H\left(X\right), \tag{54}$$

$$H(X,V\,|\,U,Y) = H(X\,|\,U,Y) + H(V\,|\,U,Y), \tag{55}$$

$$= H(X,U,Y) - H(U,Y) + H(V\,|\,U,Y), \tag{56}$$

$$= H(U,Y\,|\,X) + H(X) - H(U,Y) + H(V\,|\,U,Y), \tag{57}$$

$$= H(U\,|\,X) + H(Y\,|\,X) + H(X) - H(U,Y) + H(V\,|\,U,Y), \tag{58}$$

where (53) follows using (43), (54) follows from the chain rule for joint entropy, (55) follows using (44), (56) and (57) follow using chain rule for entropy, (58) follows using (6).

Now, it is time to sum up things. Using (53) and (54) in (51), using (58) in (52), and equating (51) and (52) yields:

$$H(V|U,Y) = H(U|X,V) + H(Y|X,V) + H(V|X) - H(U|X) - H(Y|X),$$

$$= H(U|X,V) + H(Y|X,V) + H(V|X) - H(p_1) - H(p_2), \tag{59}$$

where (59) follows using (38) and (39).

Now, we evaluate the remaining terms in (59). First, observe that, using the definition of auxiliary random variable $V$, we have

$$p(v = k|x = k) \quad = \quad 1 - p_1 - p_2 + 2p_1p_2, \tag{60}$$

$$p(v = k|x = \bar{k}) \quad = \quad p_1 + p_2 - 2p_1p_2, \tag{61}$$

where $\bar{k} \stackrel{\triangle}{=} 1 \oplus k$ for all $k \in \{0, 1\}$. Hence,

$$H(V \mid X) = H(p_1 + p_2 - 2p_1p_2). \tag{62}$$

Next, we proceed with evaluating the first term on the right hand side of (59)

$$H(U|X, V) \quad = \quad H(X, U, V) - H(X, V), \tag{63}$$

$$= \quad H(X, V|U) + H(U) - H(X, V), \tag{64}$$

$$= \quad H(X|U) + H(V|U) + H(U) - H(V|X) - H(X), \tag{65}$$

$$= \quad H(X, U) + H(V|U) - H(V|X) - H(X), \tag{66}$$

$$= \quad H(U|X) + H(V|U) - H(V|X), \tag{67}$$

$$= \quad H(p_2) + H(p_1) - H(p_1 + p_2 - 2p_1p_2), \tag{68}$$

where (63) and (64) follow using the chain rule for entropy, (65) follows using (43), (66) and (67) follow using the chain rule for entropy, (68) follows using (38), (39) and (62). Next, we proceed with evaluating the second term on the right hand side of (59)

$$H(Y|X, V) \quad = \quad H(X, Y, V) - H(X, V), \tag{69}$$

$$= \quad H(X, V|Y) + H(Y) - H(X, V), \tag{70}$$

$$= \quad H(X|Y) + H(V|Y) + H(Y) - H(V|X) - H(X), \tag{71}$$

$$= \quad H(X, Y) + H(V|Y) - H(V|X) - H(X), \tag{72}$$

$$= \quad H(Y|X) + H(V|Y) - H(V|X), \tag{73}$$

$$= \quad H(p_1) + H(p_2) - H(p_1 + p_2 - 2p_1p_2), \tag{74}$$

where (69) and (70) follow using the chain rule for entropy, (71) follows using (44), (72) and (73) follow using the chain rule for entropy, (74) follows using (38), (39) and (62).

Using (62), (68) and (74) in (59) yields

$$H(V|U, Y) = H(p_1) + H(p_2) - H(p_1 + p_2 - 2p_1p_2). \tag{75}$$

Next, using (75) in (50) yields:

$$\begin{aligned} I(U;Y) &= H(V) - H(p_1 + p_2 - 2p_1p_2), \\ &\leq 1 - H(p_1 + p_2(1 - 2p_1)), \end{aligned} \tag{76}$$

where equality in (76) is achieved if and only if $V$ is Bernoulli $1/2$, which is the case if and only if $X$ is Bernoulli $1/2$. This concludes the proof of theorem. ■

*Remark 3.6:* Straightforward algebra reveals that, the capacity of the binary symmetric case, $C(p_1, p_2) = 1 - H(p_1 + p_2 - 2p_1p_2)$, is symmetric both around the line $p_1 = 1/2$ and $p_2 = 1/2$, i.e.,

$$C(p_1, p_2) = C(1 - p_1, p_2) = C(p_1, 1 - p_2) = C(1 - p_1, 1 - p_2),$$

which is quite intuitive due to the symmetric structure of the setup. Hence, w.l.o.g., we can assume that $0 \leq p_1, p_2 \leq 1/2$.

*Remark 3.7:* Note that for $0 < p_2, p_1 < 1/2$, we have

$$p_1 \leq p_1 + p_2(1 - 2p_1) \leq (1 - p_1), \tag{77}$$

where the first inequality follows since $1 - 2p_1 \geq 0$ and the second inequality follows since

$$\begin{aligned} [p_1 + p_2(1 - 2p_1) \leq (1 - p_1)] &\iff [2p_1(1 - p_2) \leq (1 - p_2)], \\ &\iff [1/2 \geq p_1]. \end{aligned}$$

Now, since the binary entropy function is monotonic increasing (resp. decreasing) for $0 \leq p \leq 1/2$ (resp. $1/2 \leq p \leq 1$) and is symmetric around $p = 1/2$, we have

$$1 - H(p_1 + p_2(1 - 2p_1)) \leq 1 - H(p_1), \tag{78}$$

where the inequality holds with equality if and only if $p_2 = 0$ and/or $p_1 = 1/2$. Recall that, the right hand side of (78) is the Shannon-capacity for binary symmetric channel with binary alphabets. Hence, a significant consequence of (78) is that, the capacity of the binary symmetric case of asymmetric codebooks setup (cf. (40)) is strictly less than the "original Shannon type" counterpart (which is a special case of the setup at hand with $p_2 = 0$) for $0 < p_2, p_1 < 1/2$. Note that, the case of $p_1 = 1/2$ corresponds to the "information erasure" case for the communication channel; hence it is not possible to transmit any information reliably neither in our setup nor in the classical Shannon setup. Next, the case of $p_2 = 1/2$ yields the capacity of $1 - H(p_1 + p_2(1 - 2p_1))|_{p_2 = 1/2} = 1 - H(1/2) = 0$ regardless of the value of $p_1$, which is also obvious, since this case corresponds to the case of "perfectly asymmetric codebooks"; in this case, it is not possible to transmit any information due to the independence of the codewords of $\mathcal{C}_X$ and $\mathcal{C}_U$; as a result, the codewords $\{\mathbf{u}^n(W)\}_{W \in \mathcal{W}}$ and the communication channel output $\mathbf{Y}^n$ are independent.

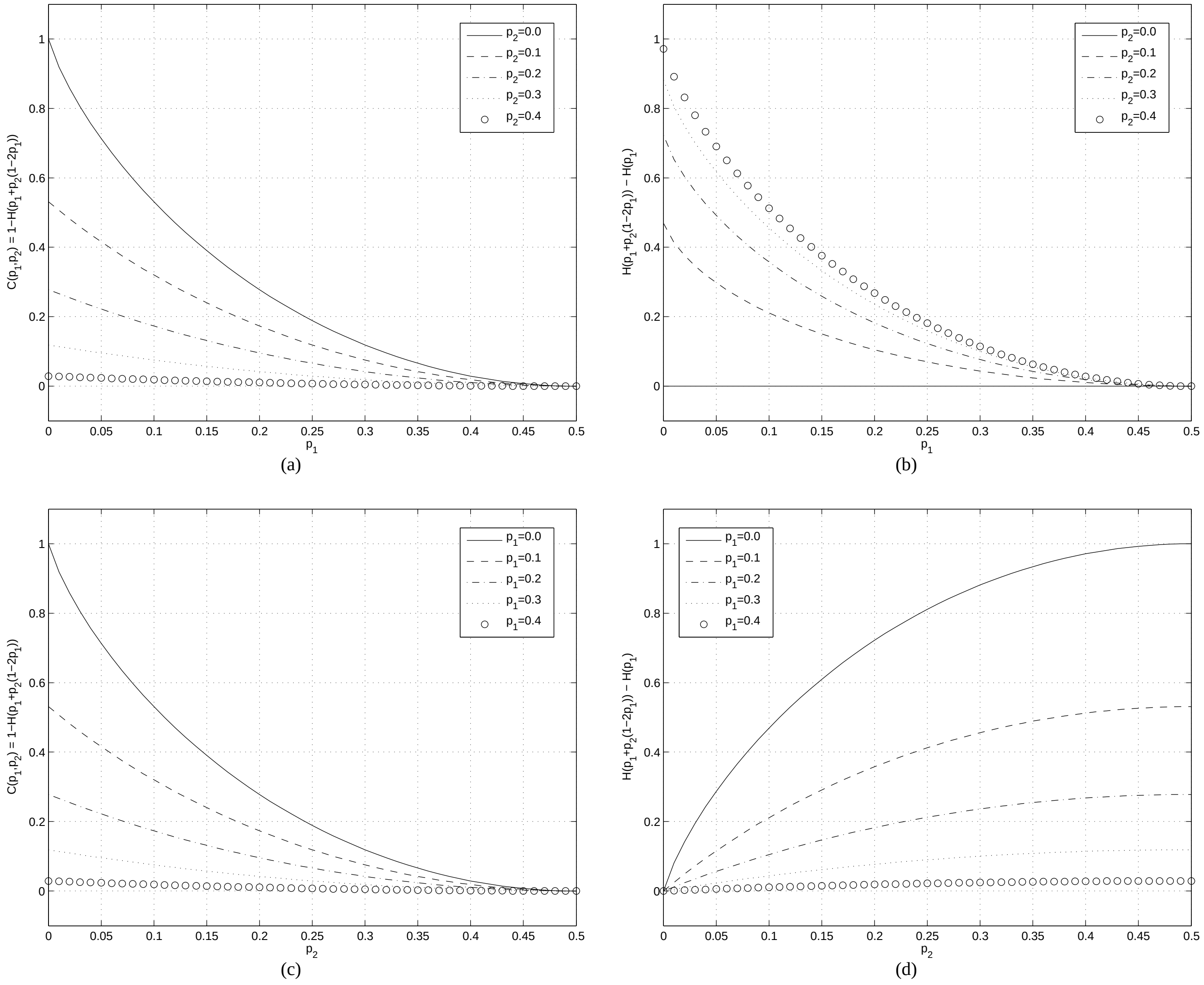

Fig. 3. Capacity results of reliable communications with asymmetric codebooks in the binary symmetric case; $p_1$ and $p_2$ denote the crossover probabilities of the communication channel and the codebook perturbation distribution, respectively. The asymmetric codebook capacity, $C\left(p_1,p_2\right)=1-H\left(p_1+p_2-2p_1p_2\right)$, is shown as a function of $p_1$ (resp. $p_2$) in panel (a) (resp. panel (c)). The gap between the aforementioned asymmetric codebook capacity and the classical Shannon capacity, $C_{Shannon}=1-H\left(p_1\right)$, which is given by $H\left(p_1+p_2-2p_1p_2\right)-H\left(p_1\right)$ is shown as a function of $p_1$ (resp. $p_2$) in panel (b) (resp. panel (d)).

*Numerical Results:* In Fig. 3, we show the numerical capacity results for the binary symmetric case. Specifically, in Fig. 3(a) and Fig. 3(c), we plot $C\left(p_1,p_2\right)=1-H\left(p_1+p_2-2p_1p_2\right)$ as functions of the communications channel crossover probability, $p_1$, and the codebook perturbation distribution crossover probability, $p_2$, respectively. From Fig. 3(a), we see that for any given $0\leq p_2\leq 1/2$, $C\left(p_1,p_2\right)$ is monotonically decreasing in $p_1$ since the communication channel becomes more noisy. Also, note that, these two figures show exactly the same behavior due to symmetry: $C\left(p_1,p_2\right)=C\left(p_2,p_1\right)$, which also implies that the same monotonic behavior of $C\left(p_1,p_2\right)$ with respect to $p_2$ for any fixed $0\leq p_1\leq 1/2$ holds. On the other hand, another quantity of interest is "the capacity loss" due to the asymmetry between the codebooks, which is given by, $C\left(p_1,p_2\right)-C_{Shannon}$, where $C_{Shannon}=C\left(p_1,p_2=0\right)=1-H\left(p_1\right)$ is the capacity of the classical binary symmetric channel setup. The capacity loss is depicted as a function of $p_1$ (resp. $p_2$) in Fig. 3(b) (resp. Fig. 3(d)). From Fig. 3(b), we see that for any given $0 \leq p_2 \leq 1/2$, the capacity loss is monotonically decreasing in $p_1$, which eventually diminishes to $0$ for the case of $p_1 = 1/2$ when no reliable transmission of information is possible both in the asymmetric and the symmetric case. From Fig. 3(d), we see that for any given $0 \leq p_1 \leq 1/2$, the capacity loss is monotonically increasing in $p_2$, which is also obvious since as $p_2$ increases, the asymmetry between the codebooks increases, thereby increasing the capacity loss.

## IV. Conclusion

In this paper, we introduced a new concept of *reliable communications with asymmetric codebooks*, which is a generalization of the classical point-to-point communication setup due to Shannon. In particular, we establish a channel coding theorem for the special case of a discrete memoryless channel with i.i.d asymmetric codebooks. We also quantify exact information capacity results for this communication system when the channel encoder codewords, decoder codewords are drawn from a binary alphabet, and the communication channel, perturbation distribution (causing the asymmetry) are analogous to the classical binary symmetric channel. Our set up is inspired by and serves as the information theoretic basis for the analysis of *robust signal hashing* where the asymmetry is due to the fact that the receiver only has access to hash values of the content owned by the transmitter. We acknowledge that the assumption of i.i.d codewords and a memoryless channel as well as codebook perturbations are not in general true for robust hashing, and aim to address those in future research. The proposed work is indeed meant to serve as a first step in information theoretic treatment of *signal hashing*.

## Acknowledgements

The authors wish to thank Süleyman Serdar Kozat of Koç University, İstanbul, Turkey, Serdar Yüksel of Queens University, Kingston, ON, Canada, Hakan Deliç of Boğaziçi University, İstanbul, Turkey, and Sviatoslav Voloshynovskiy and Oleksiy Koval of University of Geneva, Switzerland, for various helpful discussions and insightful comments, which increased the quality of the paper.

## Appendix I
## Proof of Lemma 3.1

We have

$$
\begin{aligned}
p\left(\mathbf{y}^n, \mathbf{u}^n\right) &= \sum_{\mathbf{x}^n \in \mathcal{X}^n} p\left(\mathbf{y}^n, \mathbf{u}^n, \mathbf{x}^n\right), \\
&= \sum_{\mathbf{x}^n \in \mathcal{X}^n} p\left(\mathbf{y}^n | \mathbf{u}^n, \mathbf{x}^n\right) p\left(\mathbf{u}^n | \mathbf{x}^n\right) p\left(\mathbf{x}^n\right), \\
&= \sum_{\mathbf{x}^n \in \mathcal{X}^n} p\left(\mathbf{y}^n | \mathbf{x}^n\right) p\left(\mathbf{u}^n | \mathbf{x}^n\right) p\left(\mathbf{x}^n\right), && \text{(I-1)} \\
&= \sum_{\mathbf{x}^n \in \mathcal{X}^n} \prod_{i=1}^{n} \left(p\left(y_i | x_i\right) p\left(u_i | x_i\right) p\left(x_i\right)\right), && \text{(I-2)}
\end{aligned}
$$

$$= \left( \sum_{x_1 \in \mathcal{X}} p\left(y_1|x_1\right) p\left(u_1|x_1\right) p\left(x_1\right) \right) \ldots \left( \sum_{x_n \in \mathcal{X}} p\left(y_n|x_n\right) p\left(u_n|x_n\right) p\left(x_n\right) \right)$$

$$= \prod_{i=1}^{n} \sum_{x_i \in \mathcal{X}} p\left(y_i|x_i\right) p\left(u_i|x_i\right) p\left(x_i\right),$$

$$= \prod_{i=1}^{n} p(y_i, u_i), \tag{I-3}$$

where (I-1) follows from (6), (I-2) follows using memoryless property of the communication channel and (2), and (I-3) follows by using the definition $p(y,u) = \sum_{x \in \mathcal{X}} p(y|x)p(u|x)p(x)$. Hence, the sought after result follows. □

## Appendix II
## Proof of Lemma 3.2

First, note that we have

$$\Pr\left(\mathcal{C}_X\right) = \prod_{i=1}^{n} \prod_{w=1}^{2^{nR}} p\left(x_i\left(w\right)\right), \tag{II-1}$$

$$\Pr\left(\mathcal{C}_U|\mathcal{C}_X\right) = \prod_{i=1}^{n} \prod_{w=1}^{2^{nR}} p\left(u_i\left(w\right)|x_i\left(w\right)\right), \tag{II-2}$$

where (II-1) follows from the definition of the encoder's codebook and (II-2) follows from (1) and (2).

Combining (II-1) and (II-2) yields:

$$\Pr\left(\mathcal{C}_U\right) = \sum_{\mathcal{C}_X} \prod_{i=1}^{n} \prod_{w=1}^{2^{nR}} p\left(x_i\left(w\right)\right) p\left(u_i\left(w\right)|x_i\left(w\right)\right),$$

$$= \sum_{x_1(1) \in \mathcal{X}} p\left(x_1\left(1\right)\right) p\left(u_1\left(1\right)|x_1\left(1\right)\right) \cdot \ldots \cdot \sum_{x_n(2^{nR}) \in \mathcal{X}} p\left(x_n\left(2^{nR}\right)\right) p\left(u_n\left(2^{nR}\right)|x_n\left(2^{nR}\right)\right),$$

$$= \prod_{w=1}^{2^{nR}} \prod_{i=1}^{n} \sum_{x_i(w) \in \mathcal{X}} p\left(x_i\left(w\right)\right) p\left(u_i\left(w\right)|x_i\left(w\right)\right),$$

$$= \prod_{w=1}^{2^{nR}} \prod_{i=1}^{n} p\left(u_i\left(w\right)\right), \tag{II-3}$$

where $p\left(u_i\left(w\right)\right) \triangleq \sum_{x_i(w) \in \mathcal{X}} p\left(x_i\left(w\right)\right) p\left(u_i\left(w\right)|x_i\left(w\right)\right)$ as in the statement of the lemma. Hence, (II-3) is the desired result, which concludes the proof. □

## Appendix III
## Proof of Lemma 3.3

Proof follows by contradiction: Suppose that there exists a $\left(2^{nR}, n\right)$ asymmetric code, say $\left(\mathcal{X}, \mathcal{C}_X, p\left(u|x\right), \mathcal{U}, \mathcal{C}_U\right)$, with some decoding function $g\left(\cdot\right)$ such that $\lambda^{(n)} \to 0$ and $h\left(\cdot\right)$ is not one-to-one. Next, we note the following fact: The statement of "$h\left(\cdot\right)$ is not one-to-one" equivalently means that there exists at least one set of $m \geq 2$ messages, $\tilde{\mathcal{W}} = \{\tilde{w}_i\}_{i=1}^{m} \subseteq \mathcal{W}$, such that $\forall \tilde{w}_i \in \tilde{\mathcal{W}}$, we have

$$h\left(\tilde{w}_i\right) = \mathbf{u}^n\left(\tilde{w}_i\right) = \mathbf{c}^n \tag{III-1}$$

for some constant (in $\tilde{w}_i$) length-$n$ vector $\mathbf{c}^n \in \mathcal{U}^n$. Here, let $\lambda_i$ and $\lambda^{(n)}$ denote the conditional probability of error and maximal probability of error corresponding to the code $(\mathcal{X}, \mathcal{C}_X, p(u|x), \mathcal{U}, \mathcal{C}_U)$ with the decoding function $g(\cdot)$, respectively.

Next, consider another asymmetric code, $(\mathcal{X}, \mathcal{C}_X, p(u|x), \mathcal{U}, \mathcal{C}_U)_{MAP}$, which is "derived" from $(\mathcal{X}, \mathcal{C}_X, p(u|x), \mathcal{U}, \mathcal{C}_U)$ in the following way: $(\mathcal{X}, \mathcal{C}_X, p(u|x), \mathcal{U}, \mathcal{C}_U)_{MAP}$ is the same as $(\mathcal{X}, \mathcal{C}_X, p(u|x), \mathcal{U}, \mathcal{C}_U)$, except for the decoding function; the code $(\mathcal{X}, \mathcal{C}_X, p(u|x), \mathcal{U}, \mathcal{C}_U)_{MAP}$ employs the MAP (maximum a-posterori) decoding rule [12] for the given codebooks and the communications channel, and the corresponding decoding function is denoted by $g_{MAP}(\cdot)$ (which is potentially different from the decoding function $g(\cdot)$ employed by $(\mathcal{X}, \mathcal{C}_X, p(u|x), \mathcal{U}, \mathcal{C}_U)$). Let $\lambda_{i,MAP}$ and $\lambda^{(n)}_{MAP}$ denote the conditional probability of error and maximal probability of error corresponding to the code $(\mathcal{X}, \mathcal{C}_X, p(u|x), \mathcal{U}, \mathcal{C}_U)_{MAP}$ with the decoding function $g_{MAP}(\cdot)$, respectively.

Now, recalling the definition of the MAP decoder [12], we conclude that under the assumption of uniform cost and priors[10], the MAP decoder minimizes the conditional probability of error given the codebooks and the communication channel. Therefore, we have

$$\forall i \in \mathcal{W},\ \lambda_i \geq \lambda_{i,MAP}. \tag{III-2}$$

Hence, (III-2) implies that

$$\lambda^{(n)} = \max_{i \in \mathcal{W}} \lambda_i \geq \lambda^{(n)}_{MAP} = \max_{i \in \mathcal{W}} \lambda_{i,MAP}. \tag{III-3}$$

Also, for all $\tilde{w}_i \in \tilde{\mathcal{W}}$, and for any channel input $\mathbf{x}^n$ and any channel output $\mathbf{y}^n$, note that we have

$$p\left(\mathbf{y}^n \,\middle|\, h(W = \tilde{w}_1) = \mathbf{c}^n\right) = p\left(\mathbf{y}^n \,\middle|\, h(W = \tilde{w}_2) = \mathbf{c}^n\right) = \ldots = p\left(\mathbf{y}^n \,\middle|\, h(W = \tilde{w}_m) = \mathbf{c}^n\right), \tag{III-4}$$

by using (III-1). Next, denoting the decoder output by $\hat{W}$, note that the MAP decoding rule is given by

$$\hat{W} = \arg\max_{w \in \mathcal{W}} p\left(\mathbf{y}^n \,\middle|\, \mathbf{u}^n(w)\right),$$

which is not necessarily unique in general. In particular, if we have

$$p\left(\mathbf{y}^n \,\middle|\, \mathbf{c}^n\right) = \max_{w \in \mathcal{W}} p\left(\mathbf{y}^n \,\middle|\, \mathbf{u}^n(w)\right), \tag{III-5}$$

any element $w' \in \tilde{\mathcal{W}}$ is a maximizer (cf. (III-4)). Let $\mathcal{A}(\mathbf{y}^n)$ denote some (potentially randomized) MAP decision rule if (III-5) holds. Note that, any MAP decision rule should necessarily apply some mapping $\mathcal{A}(\cdot)$, of which range is $\tilde{\mathcal{W}}$, if (III-5) holds.

Now, suppose some $w' \in \tilde{\mathcal{W}}$ has been transmitted and the $(2^{nR}, n)$ asymmetric code, $(\mathcal{X}, \mathcal{C}_X, p(u|x), \mathcal{U}, \mathcal{C}_U)_{MAP}$, with the decoding function $g_{MAP}(\cdot)$ (which incorporates some mapping $\mathcal{A} : \mathcal{Y}^n \to \tilde{\mathcal{W}}$), is applied. Then, for any $w' \in \tilde{\mathcal{W}}$, we

[10]The assumption of uniform priors follows from our problem definition; the assumption of uniform costs follows from the fact that we are interested in the case which minimizes the probability of error, i.e., Minimum Probability of Error (MPE) rule, which is equivalent to the MAP rule under the specified assumptions.

have

$$1-\lambda_{w',MAP} = \Pr\left[\hat{W}=w' \,\middle|\, W=w'\right], \tag{III-6}$$

$$\leq \Pr\left[\hat{W}=w' \,\middle|\, p\left(\mathbf{y}^n \mid \mathbf{c}^n\right)=\max_{w\in\mathcal{W}} p\left(\mathbf{y}^n|\mathbf{u}^n\left(w\right)\right),\ W=w'\right] \tag{III-7}$$

$$= \Pr\left[\hat{W}=w' \,\middle|\, p\left(\mathbf{y}^n \mid \mathbf{c}^n\right)=\max_{w\in\mathcal{W}} p\left(\mathbf{y}^n|\mathbf{u}^n\left(w\right)\right)\right] \tag{III-8}$$

$$= \Pr\left[\mathcal{A}\left(\mathbf{y}^n\right)=w'\right], \tag{III-9}$$

where (III-6) follows from the definition of $\{\lambda_{w',MAP}\}$, (III-7) follows since the event $p\left(\mathbf{y}^n \mid \mathbf{c}^n\right)=\max_{w\in\mathcal{W}} p\left(\mathbf{y}^n|\mathbf{u}^n\left(w\right)\right)$ is a necessary condition for the event $\hat{W}=w'$ for the case of MAP decoding, (III-8) follows since $\max_{w\in\mathcal{W}} p\left(\mathbf{y}^n|\mathbf{u}^n\left(w\right)\right)$ is a sufficient statistic for MAP decoding, (III-9) follows from the definition of the MAP decoding rule $g_{MAP}\left(\cdot\right)$ and the utilized mapping $\mathcal{A}\left(\cdot\right)$. Hence, using (III-9), for any $w'\in\tilde{\mathcal{W}}$ we have

$$\lambda_{w',MAP}\geq 1-\Pr\left[\mathcal{A}\left(\mathbf{y}^n\right)=w'\right],$$

which, in turn, implies

$$\max_{w'\in\tilde{\mathcal{W}}}\lambda_{w',MAP}\geq 1-\min_{w'\in\tilde{\mathcal{W}}}\Pr\left[\mathcal{A}\left(\mathbf{y}^n\right)=w'\right]. \tag{III-10}$$

Next, upon defining $q\left(w'\right)\triangleq\Pr\left[\mathcal{A}\left(\mathbf{y}^n\right)=w'\right]$, we note that $\{q\left(w'\right)\}_{w'\in c\tilde{W}}$ is a valid p.m.f. over the discrete finite set $\tilde{\mathcal{W}}$, i.e., $\forall w'\in\tilde{\mathcal{W}}$, $q\left(w'\right)\in[0,1]$ and $\sum_{w'\in\tilde{\mathcal{W}}}q\left(w'\right)=1$. Therefore, we clearly have

$$\max_{\mathcal{A}(\cdot)}\min_{w'\in\tilde{\mathcal{W}}}q\left(w'\right)=\frac{1}{\left|\tilde{\mathcal{W}}\right|}=\frac{1}{m},$$

which implies

$$\min_{w'\in\tilde{\mathcal{W}}}\Pr\left[\mathcal{A}\left(\mathbf{y}^n\right)=w'\right]\leq\frac{1}{m}. \tag{III-11}$$

Thus, using (III-11) in (III-10)

$$\max_{w'\in\tilde{\mathcal{W}}}\lambda_{w',MAP}\geq 1-\frac{1}{m}. \tag{III-12}$$

Consequently, we have

$$\lambda^{(n)}\geq\lambda^{(n)}_{MAP}\geq\max_{w'\in\tilde{\mathcal{W}}}\lambda_{w',MAP}\geq 1-\frac{1}{m}\geq\frac{1}{2}>0, \tag{III-13}$$

where the first inequality follows from (III-3), the second inequality follows since $\tilde{\mathcal{W}}\subseteq\mathcal{W}$, the third inequality follows from (III-12), the fourth inequality follows since $m\geq 2$. Hence, the promised contradiction follows from (III-13). ☐

## APPENDIX IV
## PROOF OF LEMMA 3.4

First of all, observe that we have $U = V \oplus Z_1$, and $Y = V \oplus Z_2$ (cf. (38) and (39))

$$\Pr\left(U = u | V = v\right) = \Pr\left(Z_1 = u \oplus v\right), \tag{IV-1}$$

$$\Pr\left(Y = y | V = v\right) = \Pr\left(Z_2 = y \oplus v\right), \tag{IV-2}$$

for all $u, v, y \in \{0, 1\}$. Furthermore, we also have

$$\Pr\left(U = u, Y = y | V = v\right) = \Pr\left(Z_1 = u \oplus v, Z_2 = y \oplus v\right), \tag{IV-3}$$

$$= \Pr\left(Z_1 = u \oplus v\right) \Pr\left(Z_2 = y \oplus v\right), \tag{IV-4}$$

for all $u, v, y \in \{0, 1\}$, where (IV-3) follows using (38) and (39), and (IV-4) follows since $Z_1$ and $Z_2$ are independent. Combining (IV-1), (IV-2) and (IV-4), we conclude that $p(u, y|v) = p(u|v)p(y|v)$, which is the sought-after result. □